\documentclass[a4paper,12pt,times,oneside,onecolumn,preprint]{elsarticle}
\usepackage[dvipsnames, svgnames, x11names]{xcolor}
 \usepackage{amsthm}
 \usepackage{graphics}
 \usepackage{array}
\usepackage{booktabs}

\usepackage{graphicx,subfigure}
 \usepackage{graphicx}
 \usepackage{epsfig}
  \usepackage{paralist}
  \usepackage{epstopdf}
   \usepackage[colorlinks=true]{hyperref}

\usepackage{epsfig,graphicx,amssymb,amsmath}

\allowdisplaybreaks[4]

\usepackage{tikz}
\usetikzlibrary{arrows,automata}
\usepackage[latin1]{inputenc}
\usepackage{verbatim}
\usepackage{tikz}
\usepackage{tikz}
\usetikzlibrary{arrows,calc,shapes,decorations.pathreplacing}

\usepackage{colortbl}
\usepackage{color}
\usepackage{tabularx}
\usepackage{booktabs}
\usepackage{caption}
\usetikzlibrary{calc, positioning}
\pgfdeclarelayer{background}
\pgfdeclarelayer{foreground}
\pgfsetlayers{background,main,foreground}
\usepackage{ifthen}
\usepackage{animate}
\usepackage{graphicx,subfigure}
\usepackage{float}
\usepackage{epsfig,graphicx,amssymb,amsmath}
\usetikzlibrary{shapes.arrows}
\usepackage{textcomp,booktabs}

\usepackage{empheq}
\usepackage{amsfonts}
\usepackage{makecell}
\usepackage{amsmath}
\usepackage{epstopdf}
\usepackage{tikz}
\journal{ Applied Mathematics and Computation }
\begin{document}
\begin{frontmatter}
\title{Effects of Dynamic-Win-Stay-Lose-Learn model with voluntary participation in social dilemma}
\author[]{Zhenyu Shi \textsuperscript{1,2,3,4}}
\author[]{Wei Wei \textsuperscript{1,2,3,4,*}}
\author[]{Xiangnan Feng \textsuperscript{1,2,3,4}}
\author[]{Ruizhi Zhang \textsuperscript{1,2,3,4}}
\author[]{Zhiming Zheng \textsuperscript{1,2,3,4}}

\ead{}
\cortext[]{Corresponding authors: Wei Wei}
\address{1.School of Mathematical Sciences, Beihang University, Beijing, China

2.Key Laboratory of Mathematics Informatics Behavioral Semantics, Ministry of Education, China

3.Beijing Advanced Innovation Center for Big Data and Brain Computing, Beihang

4.Peng Cheng Laboratory, Shenzhen, Guangdong, China

}
\begin{abstract}
In recent years, Win-Stay-Lose-Learn rule has attracted wide attention as an effective strategy updating rule, and voluntary participation is proposed by introducing a third strategy in Prisoner's dilemma game.
Some researches show that combining Win-Stay-Lose-Learn rule with voluntary participation could promote cooperation more significantly under moderate temptation values,
however, cooperators' survival under high aspiration levels and high temptation values is still a challenging problem.
In this paper, inspired by Achievement Motivation Theory, a Dynamic-Win-Stay-Lose-Learn rule with voluntary participation is investigated, where a dynamic aspiration process is introduced to describe the co-evolution of individuals' strategies and aspirations.
It is found that cooperation is extremely promoted and defection is almost extinct in our model, even when the initial aspiration levels and temptation values are high.
The combination of dynamic aspiration and voluntary participation play an active role since loners could survive under high initial aspiration levels and they will expand stably because of their fixed payoffs.
The robustness of our model is also discussed and some adverse structures are found which should be alerted in the evolutionary process.
Our work provides a more realistic model and shows that cooperators may prevail defectors in an unfavorable initial environment.
\end{abstract}

\begin{keyword}
Prisoner's dilemma game, Dynamic aspiration, Cooperation, Spatial evolutionary game
\end{keyword}
\end{frontmatter}

\section{Introduction}
Though cooperative behavior is ubiquitous in biological, economic and social systems \cite{Myerson1991Game, schuster2004Game, gibbons1992Game}, how to explain its emergence and stability is still a valuable and challengeable problem in related fields \cite{axelrod1981evolution}.
Prisoner's dilemma (PD) game is a representative model to describe the social dilemmas among selfish individuals.
In a typical PD game, two players choose cooperation or defection simultaneously without communication.
One will always get a higher payoff if it chooses defection whichever its opponent chooses, but if both of them choose defection, their payoffs are lower than those of both choosing cooperation, which leads to a conflict between individual rationality and collective rationality.
Evolutionary game theory has been thought of as a powerful mathematical framework to reveal the mechanisms for cooperation phenomenon in the competitive environment.
In \cite{nowak2006five}, Nowak found five typical mechanisms which are conducive to the existence of cooperation: kin selection, direct and indirect reciprocity, network reciprocity, and group selection, among which network reciprocity has received the most widespread attention.
Series of researches focus on different kinds of network structures, such as lattice \cite{nowak1992spatial, Szab1998lattice}, small world networks \cite{2005smallworld, 2007smallworld} and scale-free networks \cite{2007scalefree, Du2009scalefree}.
Besides, many social mechanisms have been proved to enhance cooperation, such as punishment \cite{herrmann2008punishment, helbing2010punishment, chen2014punishment, chen2015punishment}, game organizers \cite{szolnoki2015conformist, szolnoki2016conformist}, compassion \cite{2018compassion}, memory effects \cite{2006Memory, 2008Memory}, and so on.

In recent years, aspiration-based strategy updating rules have got more and more attention from researchers \cite{Yang2012aspiration, Wu2018aspiration, Chu2019aspiration, Zhang2019aspiration, posch1999dynamic, amaral2016dynamic, arefin2020dynamic, 2020dynamic, shi2021dynamic}.
A representative model is that compared to maximize their payoffs, individuals usually tend to keep their strategies when they feel satisfied, otherwise they try to learn what others do, which is called Win-Stay-Lose-Learn strategy updating rule and there have been some related studies in recent years \cite{liu2012win, chu2017win, fu2018win}.
In most of the above research, aspiration is a fixed value which is set for all individuals before the evolution process begins\cite{Yang2012aspiration, Wu2018aspiration, Chu2019aspiration, Zhang2019aspiration, liu2012win, chu2017win, fu2018win}.
In fact, dynamic aspiration models meet the actual situation better and there are also some related researches \cite{posch1999dynamic, amaral2016dynamic, arefin2020dynamic, 2020dynamic, shi2021dynamic}.
But the mechanism how dynamic aspiration models impact the evolution process is still to be resolved.
Our work focuses on what role dynamic aspiration models play to promote cooperation, which is compared to fixed aspiration model.

Based on traditional PD games, it is found that introducing a new strategy to PD games may promote cooperation, such as tit-for-tat \cite{Szolnoki2009tit}, punishment \cite{2017punishment} and voluntary participation \cite{2016lone}.
Voluntary participation is one of the new strategies, which means an individual may choose to abstain the PD game and get a low but guaranteed payoff.
It appropriately describes the phenomenon that some individuals may choose to reduce interactions in social dilemmas, and causes a three-strategies evolution process.

Although Win-Stay-Lose-Learn rule and voluntary participation could promote cooperation under moderate temptation values, cooperators' survival under high aspiration levels and high temptation values is still a challenging problem.
In this paper, combining both of their advantages, a Dynamic-Win-Stay-Lose-Learn rule is proposed in the Optional Prisoner's Dilemma game.
The model aims at promoting cooperation obviously with the appropriate initial structure, even when the temptation value $T$ is large.

In the remainder of our paper, firstly we introduce our Dynamic-Win-Stay-Lose-Learn strategy updating rule with voluntary participation in the \emph{Model} section.
Then the main valuable phenomena is shown based on the results of Monte Carlo simulation, and more detailed analysis about why loners could promote cooperation under dynamic aspiration model and what are the necessary conditions for promoting cooperation is discussed, which is divide into two different parameter regions in the \emph{Results} section.
Finally the main conclusion and innovation of our work is summarized in the \emph{Conclusion} section.

\section{Model}\label{section 3}
Our work considers PD games with voluntary participation in which three strategies are included: cooperation($\mathcal{C}$), defection($\mathcal{D}$) and lone($\mathcal{L}$), which is also called Optional Prisoner's Dilemma(OPD) game.
Individuals distribute in an $L \times L$ square lattice with periodic boundary conditions, in this paper, $L$ is set to 100, and each of them will only interact with its four direct neighbors.
The strategy and aspiration of an individual $i$ are denoted as $s_i$ and $A_i$ respectively.
In the evolutionary process, all individuals update their strategies and aspirations synchronously by discrete time steps.
One complete step performs according to following rules:

(a)\textbf{Rule of game:} Each individual $i$ plays OPD games with its four direct neighbors to get a payoff $P_i = \sum\limits_{j\in {\Omega_i}} P_{ij}$, where $\Omega_i$ represents all direct neighbors of individual $i$.
$P_{ij}$ represents $i$'s payoff for playing an OPD game with $j$, which could be got by Table 1.
They will receive the reward $R$ or punishment $P$ if they both choose to cooperate or defect.
If one of them chooses $\mathcal{C}$ but the other one chooses $\mathcal{D}$, the former will receive the sucker's payoff $S$ and the the latter will receive the temptation value $T$.
If at least one of them choose $\mathcal{L}$, both of them will get the loner's payoff $l$.
In OPD games, $\emph{S}\textless \emph{P}\textless \emph{R}\textless \emph{T}$ and $0 \textless l \textless 1$ should be meet.

\begin{table}[h]
\caption{Payoff matrix of OPD games.}
\begin{center}
\begin{tabular}{|c|c|c|c|}
     \hline
     &\textbf{$\mathcal{C}$}&\textbf{$\mathcal{D}$}&\textbf{$\mathcal{L}$}  \\
     \hline
     \textbf{$\quad$ $\mathcal{C}$$\quad$}&$\quad R \quad $&$\quad S \quad $ &$\quad l \quad $\\
     \hline
     \textbf{$\quad$ $\mathcal{D}$$\quad$}&$T$&$P$ &$\quad l \quad $\\
     \hline
     \textbf{$\quad$ $\mathcal{L}$$\quad$}&$l$&$l$ &$\quad l \quad $\\
     \hline
\end{tabular}
\end{center}
\end{table}
In this paper, parameters are set as boundary game: $\emph{R}=1$, $\emph{P}=\emph{S}=0$ and $\emph{T}=b$ \cite{nowak1992spatial, liu2012win}.
And $l$ is set to 0.3 as what \cite{chu2017win} does.


(b)\textbf{Rule of strategy's update:} Each individual $i$ compares $P_i$ with $A_i$.
If $P_i \geq A_i$, $i$ will keep its strategy.
If $P_i \textless A_i$, $i$ will select one of its direct neighbors $j$ at random and imitate $j$'s strategy with the Fermi updating rule:

\begin{equation}
W_{ij}=\frac{1}{1+\exp[(P_i-P_j)/K]},
\end{equation}
where $K$ represents the amplitude of noise \cite{szabo1998noise} and is set to 0.1 in our model to characterize appropriate randomness \cite{szabo2005noise, perc2006noise}.

(c)\textbf{Rule of aspiration's update:} Each individual $i$ updates its aspiration according to the difference between $P_i$ with $A_i$ with the linear updating rule:

\begin{equation}
A_i(t+1) = A_i(t) + a * (P_i(t) - A_i(t)),
\end{equation}
where $A_i(0)=A$ is the initial aspiration for all individuals.
The dynamic-aspiration process is represented by a linear updating rule, where $a\in[0,1]$ quantifies the evolution rate of aspiration \cite{shi2021dynamic}.
When $a=0$, the model is reduced to the fixed aspiration model.
The upper bound of $a=1$ ensures that one's aspiration will be close to but not over payoff.
In general, an individual's aspiration might not be updated drastically, so $a$ should set to a small value.
In this paper, we set $a=0.05$.


To ensure the network to be stable, above step will carry out 100,000 times repeatedly in a simulation.
The final fractions of the three strategies denoting as $r_\mathcal{C}$, $r_\mathcal{D}$ and $r_\mathcal{L}$ are calculated by the average of the last 1,000 steps.
For each pair of parameters, 20 independent simulations are performed to make the results more accurate.

\section{Results}\label{section 4}

\subsection{Overview}
\begin{figure}[!htpb]

\centering

\includegraphics[width=7in]{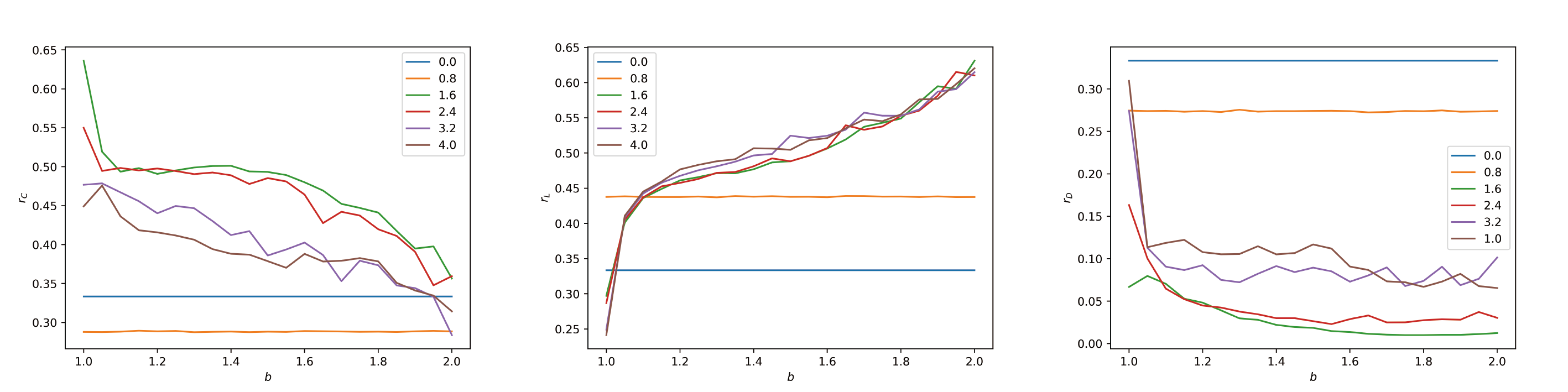}

\caption{Average fractions of cooperators, loners and defectors in the stable state in dependence on $b$ at different values of the $A$, from left to right respectively.}\label{1}

\end{figure}

To begin our discussion, the result of randomly initialized network is shown, where all of the three strategies occupy one third of the network with random settings.
The fractions of three strategies are donated as $r_{\mathcal{C}0}$, $r_{\mathcal{D}0}$ and $r_{\mathcal{L}0}$.
Besides, for all individuals, the initial aspiration $A$ is the same value.
Figure \ref{1} shows $r_\mathcal{C}$, $r_\mathcal{L}$ and $r_\mathcal{D}$ as a function depending on $b$ for different values of $A$.
It could be found that there are significant differences between the results for different values of $A$.
Two different phases could be easily observed, which are called \emph{Stable Coexistence} and \emph{Defection Suppression} respectively.
When $A \leq 1.2$, cooperators, defectors and loners could coexist with moderate fractions respectively.
For instance, $r_\mathcal{C}=r_\mathcal{L}=r_\mathcal{D}=0.33$ when $A=0$, $r_\mathcal{C}=0.29$, $r_\mathcal{L}=0.44$ and $r_\mathcal{D}=0.27$ when $A=0.8$ which are independent with the value of $b$.
When $A \textgreater 1.2$, it can be observed that $r_\mathcal{C}$ decreases (monotonously) with the increase of the value of $b$ and $r_\mathcal{L}$ is the opposite.
More importantly, $r_\mathcal{D}$ always keeps a low level which is related to the value of $A$.
Compared to the results shown in \cite{chu2017win}, dynamic aspirations model with voluntary participation plays an important role to promote cooperation, especially when $b$ is close to 2.0.

\subsection{Stable Coexistence($A \leq 1.2$)}
\begin{figure}[!htpb]

\centering

\includegraphics[width=5in]{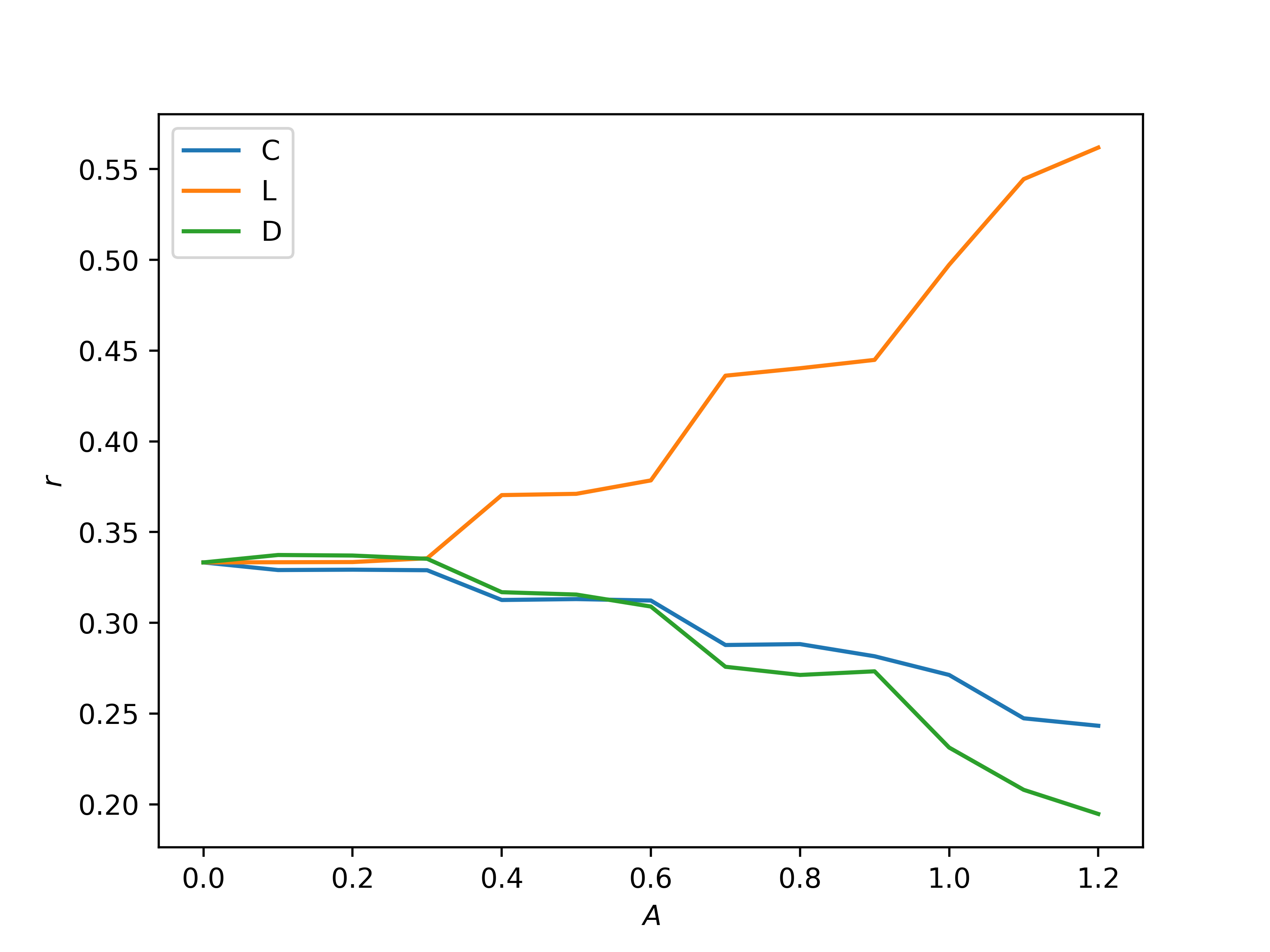}

\caption{Average fractions of cooperators, loners and defectors in the stable state in dependence on $A$ when $b=1.6$.}\label{2}

\end{figure}

For small values of $A$, all of three strategies could survive with moderate fractions.
Figure \ref{2} presents the fraction $r_\mathcal{C}$ of cooperators, the fraction $r_\mathcal{L}$ of loners and the fraction $r_\mathcal{D}$ of defectors when stable as a function of $A$ when $b=1.6$.
It is shown that with $A$ increasing, $r_\mathcal{L}$ increases but $r_\mathcal{C}$ and $r_\mathcal{D}$ decrease.
Besides, there are some apparent  discontinuous transitions which are $A=$0.3, 0.6, 0.9 and 1.0.
These transitions can be explained as follows.
In our model, the payoff of a loner is always 1.2 no matter what strategies its neighbors have.
When $A \leq 1.2$, a loner's payoff is higher than its aspiration so it is always satisfied and never changes its strategy.
As for a cooperator or a defector, their payoff can be written as $n_\mathcal{C}+0.3n_\mathcal{L}$ and $bn_\mathcal{C}+0.3n_\mathcal{L}$ respectively, where $n_\mathcal{C}$, $n_\mathcal{L}$ and $n_\mathcal{D}$ represent the number of a node's $\mathcal{C}$ neighbors, $\mathcal{L}$ neighbors and $\mathcal{D}$ neighbors and they should meet $n_\mathcal{C}+n_\mathcal{L}+n_\mathcal{D}=4$.
There might be three different states:

\begin{itemize}
\item
When $n_\mathcal{C} \textgreater 1$, the node's payoff is higher than 2.0 and always satisfied.

\item
When $n_\mathcal{C}=1$, the node might be dissatisfied only when $n_\mathcal{L}=0$. Under this state, the payoffs of a cooperator and a defector are 1.0 and $b$ respectively.

\item
When $n_\mathcal{C}=0$, the node's payoff can only be 0, 0.3, 0.6, 0.9 or 1.2.
\end{itemize}

\begin{figure}[!htpb]

\centering

\includegraphics[width=5in]{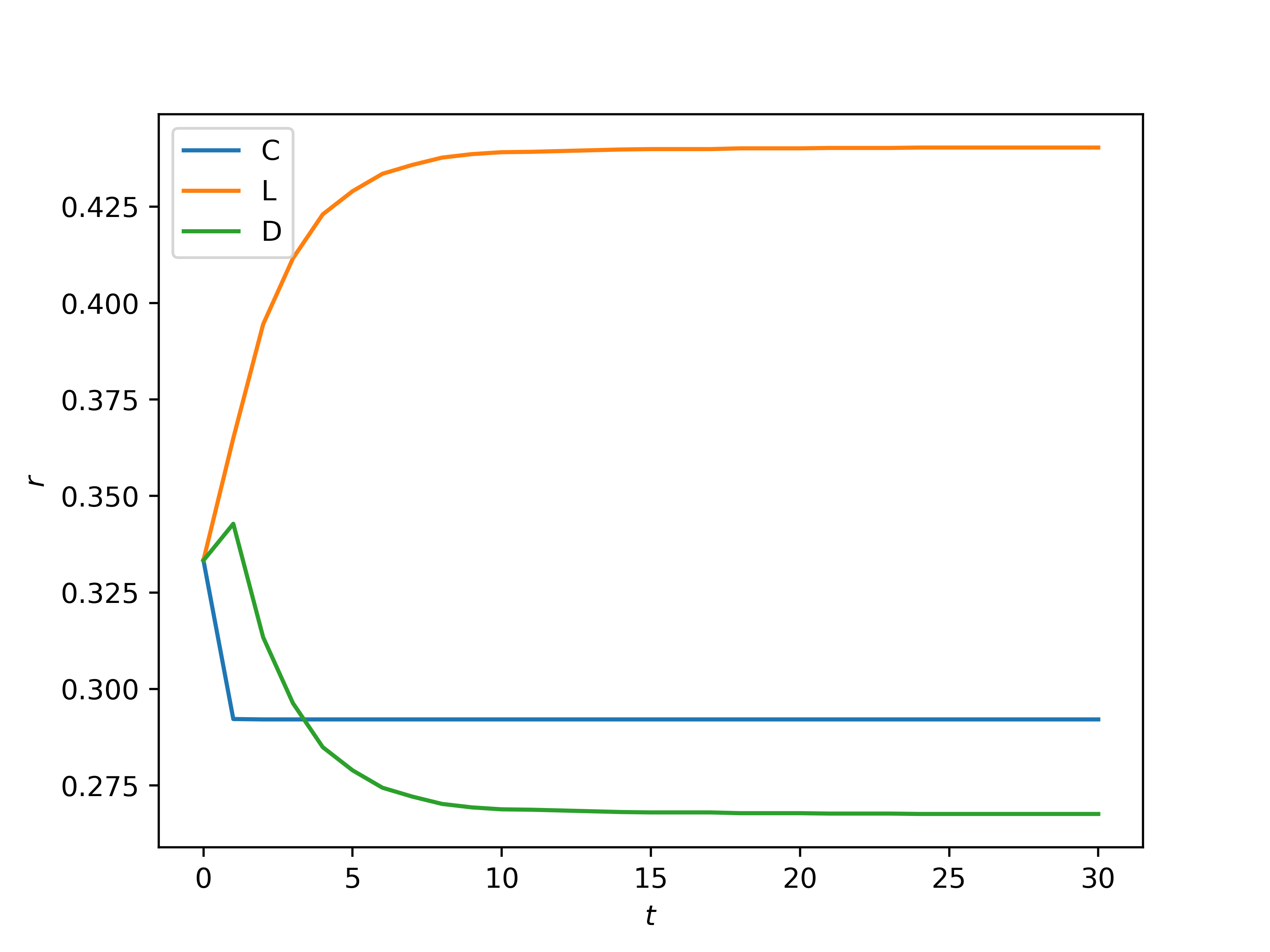}

\caption{Average fractions of cooperators, loners and defectors as a function of step $t$ when $A=0.8$ and $b=1.9$.}\label{3}

\end{figure}

As mentioned above, the possible value of a node's payoff is in \{0.3, 0.6, 0.9, 1.0\}, which are consistent with the points that discontinuous transitions happen.
When a cooperator or a defector is dissatisfied, it will change its strategy by imitating its neighbors.
Furthermore, it might become satisfied only when it evolves into a loner.
On the contrary, a loner will never change its strategy.
So it is observed that $r_\mathcal{L}$ will be higher than 0.33 (the initial fraction of $\mathcal{L}$) when stable.
The higher $A$ is, the higher $r_\mathcal{L}$ will be when stable because more cooperators and defectors will be dissatisfied and evolve into loners finally, which is independent with the value of $b$.
Figure \ref{3} presents $r_\mathcal{C}$, $r_\mathcal{L}$ and $r_\mathcal{D}$ with the time-evolution when $A=0.8$ and $b=1.9$.
Some cooperators and defectors evolve into loners quickly then the network is stable even though under high value of $b$.
Loners play an important role that they will never change their strategies if they are satisfied initially.

Besides, the theoretically values of $r_\mathcal{C}$ for different $A$ could be calculated by the formula:

\begin{equation}
r_\mathcal{C}=r_{\mathcal{C}0} * (1 - \sum_{n=0}^k C_{4}^{k}  r_{\mathcal{L}0}^k  r_{\mathcal{D}0}^{4-k}),
\end{equation}
where $k=\lceil A/0.3 \rceil-1$.

For instance, when $A=0.3$ ($k=0$), a cooperator will be dissatisfied only when all its four neighbors are defectors, with probability of $r_\mathcal{L}^4$.
So $r_\mathcal{C}$ could be calculated as:

\begin{equation}
r_\mathcal{C}=r_{\mathcal{C}0} * (1 - r_{\mathcal{D}0}^4) \approx 0.3291.
\end{equation}
which is consistent with the Monte Carlo simulation result shown in Figure \ref{2}.

\subsection{Defection Suppression($A \textgreater 1.2$)}
\begin{figure}[!htpb]

\centering

\includegraphics[width=5in]{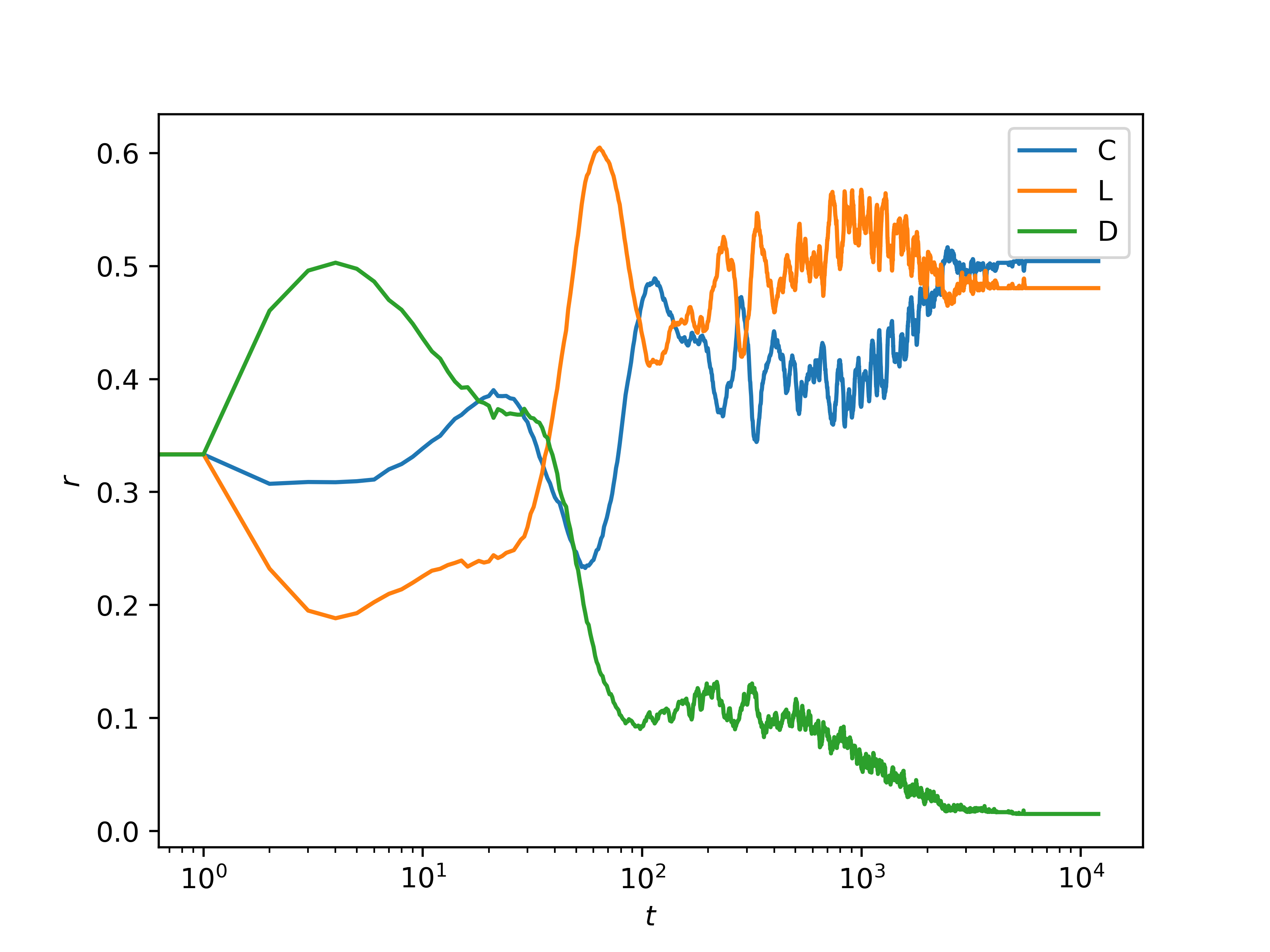}

\caption{Average fractions of cooperators, loners and defectors as a function of step $t$ when $A=1.6$ and $b=1.6$.}\label{4}

\end{figure}

When $A$ is large, cooperators and loners will expand and coexist while defectors' survival is greatly suppressed.
Figure \ref{4} presents $r_\mathcal{C}$, $r_\mathcal{L}$ and $r_\mathcal{D}$ with the time-evolution when $A=1.6$ and $b=1.6$.
It could be easily observed that there are four obvious phases:
\begin{itemize}
\item
At first, because the initial aspiration of an individual is 1.6 and a loner's payoff is always 1.2, all the loners are dissatisfied and try to change their strategies.
On the contrary, most of the cooperators and defectors are satisfied.
Besides, defectors's payoffs are higher than cooperators' on average.
So $r_\mathcal{L}$ decreases fast and defectors expand transitorily.

\item
With $t$ growing, loners' aspirations become lower than 1.2, so they get satisfied and never change their strategies any more.
Cooperators and defectors with low payoffs will try to evolve into loners and become satisfied, so loners could expand stably.
At the same time, cooperators form some clusters gradually, which is like to the so-called END period \cite{ogasawara2014end, kabir2018end}.

\item
With $t$ further growing, cooperators which still survive have formed some clusters.
Dissatisfied defectors and loners neighboring with these clusters will evolve into cooperators and cause the chain phenomenon, which causes cooperators' expansion and it is called EXP period \cite{ogasawara2014end, kabir2018end}.
$r_\mathcal{C}$ increases while $r_\mathcal{D}$ and $r_\mathcal{L}$ decrease.

\item
Finally, all of the three strategies have formed some stable clusters which will never evolve any more, while there are some regions where three strategies mix well.
In these regions, because of the cyclic dominance, Rock-Scissors-Paper-type cycles occur in three strategies.
Since loners and cooperators' clusters are easier to expand, it could be found that $r_\mathcal{C}$ and $r_\mathcal{L}$ increase while $r_\mathcal{D}$ decreases concussively.
The whole network will be stable at about $t=2000$.

\end{itemize}

\begin{figure}[!htpb]

\centering

\includegraphics[width=6in]{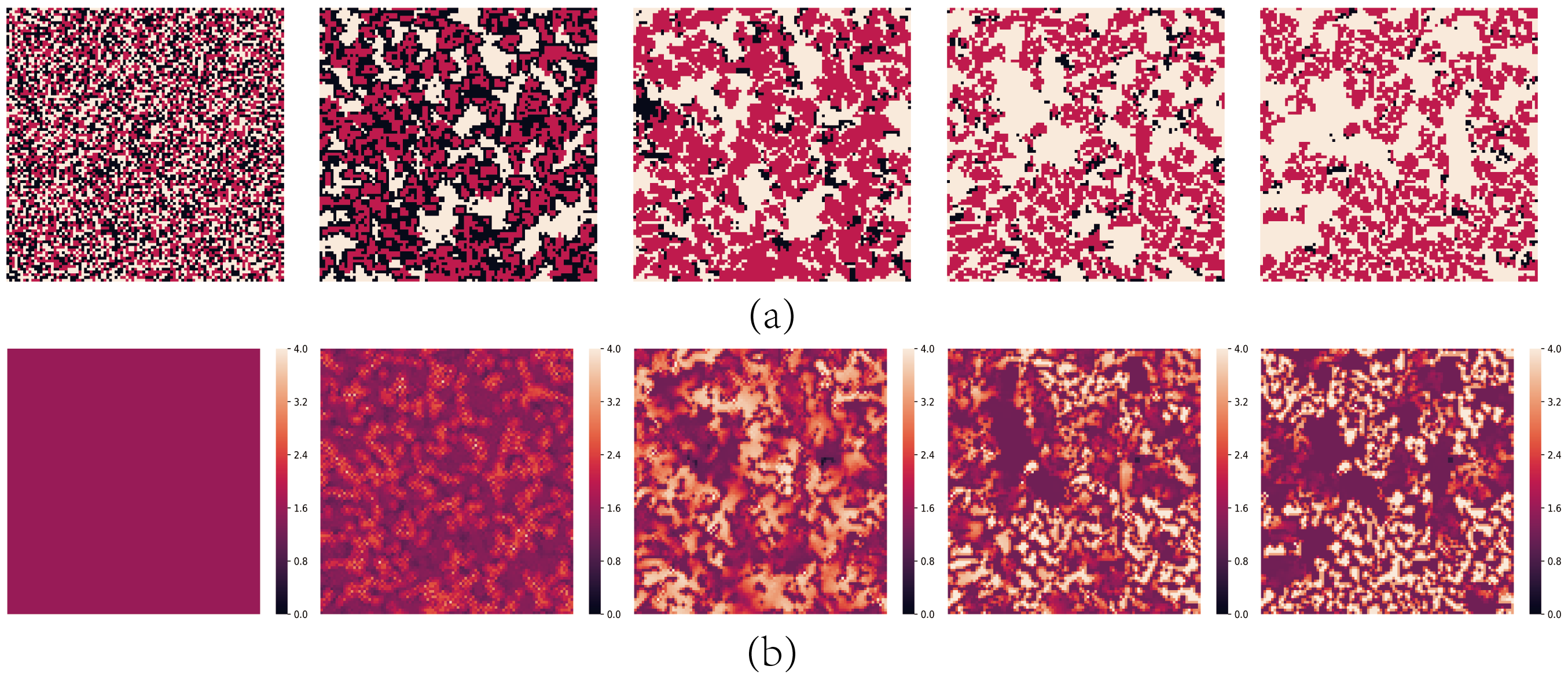}

\caption{(a) shows characteristic snapshots of cooperators(red), loners(white) and defectors(black) with time growing. (b) shows the heat map of aspiration distribution with time growing. The steps of them are $t$=0, 10, 100, 1000 and 2000 from left to right respectively. The results were got when $A=1.6$ and $b=1.6$.}\label{5}

\end{figure}

In order to further discuss how cooperators, loners and defectors and their aspiration levels distribute in the network, Figure \ref{5} represents snapshots of strategies and aspirations for $A=1.6$ and $b=1.6$.
As it is shown, cooperators and loners are separated by defectors at first and all of them are located in small-scale clusters.
When the network is stable, loners have formed several large-scale clusters while cooperators are still located in many small-scale clusters.
It is because that when neighboring directly, loners are dominant over defectors while defectors are dominant over cooperators.

\begin{figure}[!htpb]

\centering

\includegraphics[width=6in]{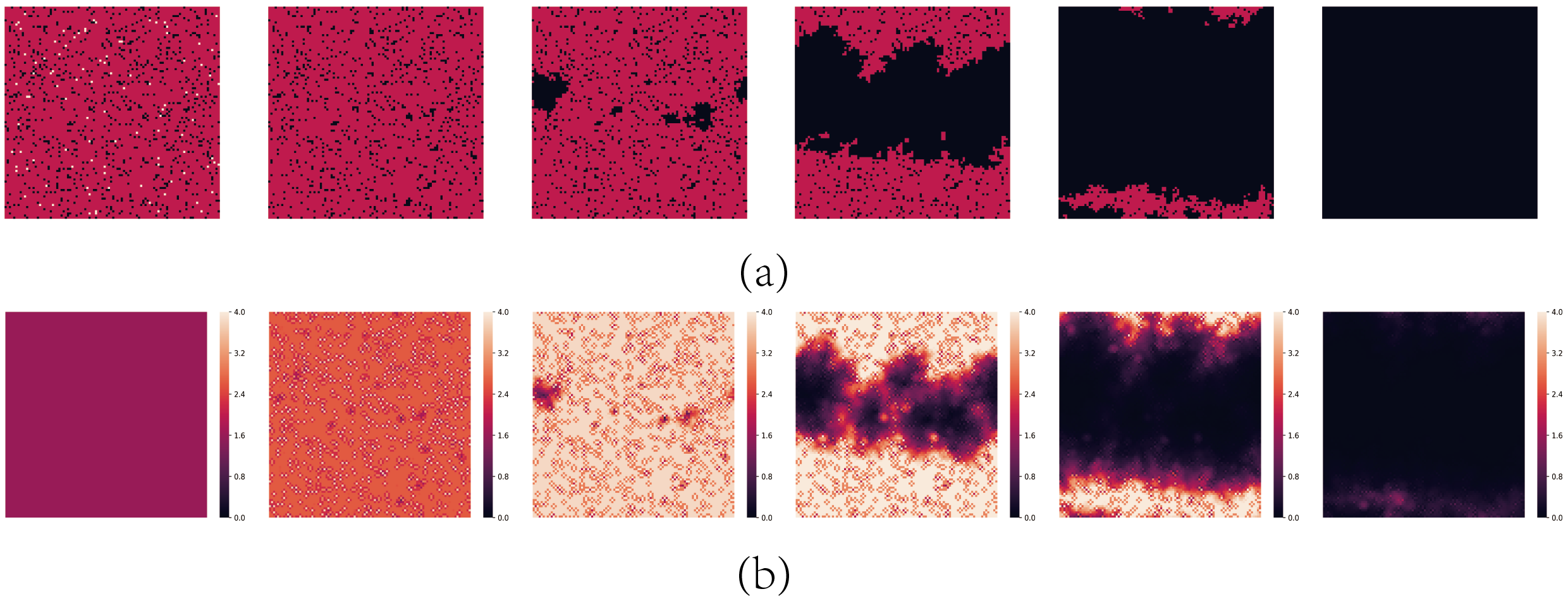}

\caption{(a) shows characteristic snapshots of cooperators(red), loners(white) and defectors(black) with time growing. (b) shows the heat map of aspiration distribution with time growing. The steps of them are $t$=0, 10, 50, 100, 150 and 200 from left to right respectively. The results were got when $A=1.6$ and $b=1.6$ with $r_{\mathcal{C}0}=0.9$, $r_{\mathcal{D}0}=0.09$ and $r_{\mathcal{L}0}=0.01$ initially.}\label{6}

\end{figure}

According to the above analysis, voluntary participation plays an important role to promote cooperation because of loner's fixed payoff. When loners get satisfied, they are certain to survive and expand.
But it should be noticed that loners might be extinct before part of them get satisfied.
As Figure \ref{6} showing, loners will be extinct soon when the value of  $r_{\mathcal{L}0}$ is too small.
Then the network degenerates to the two-strategies condition where defectors could expand to the whole network under such parameters. In this case, cooperators couldn't survive with the help of loners.

\begin{figure}[!htpb]

\centering

\includegraphics[width=5in]{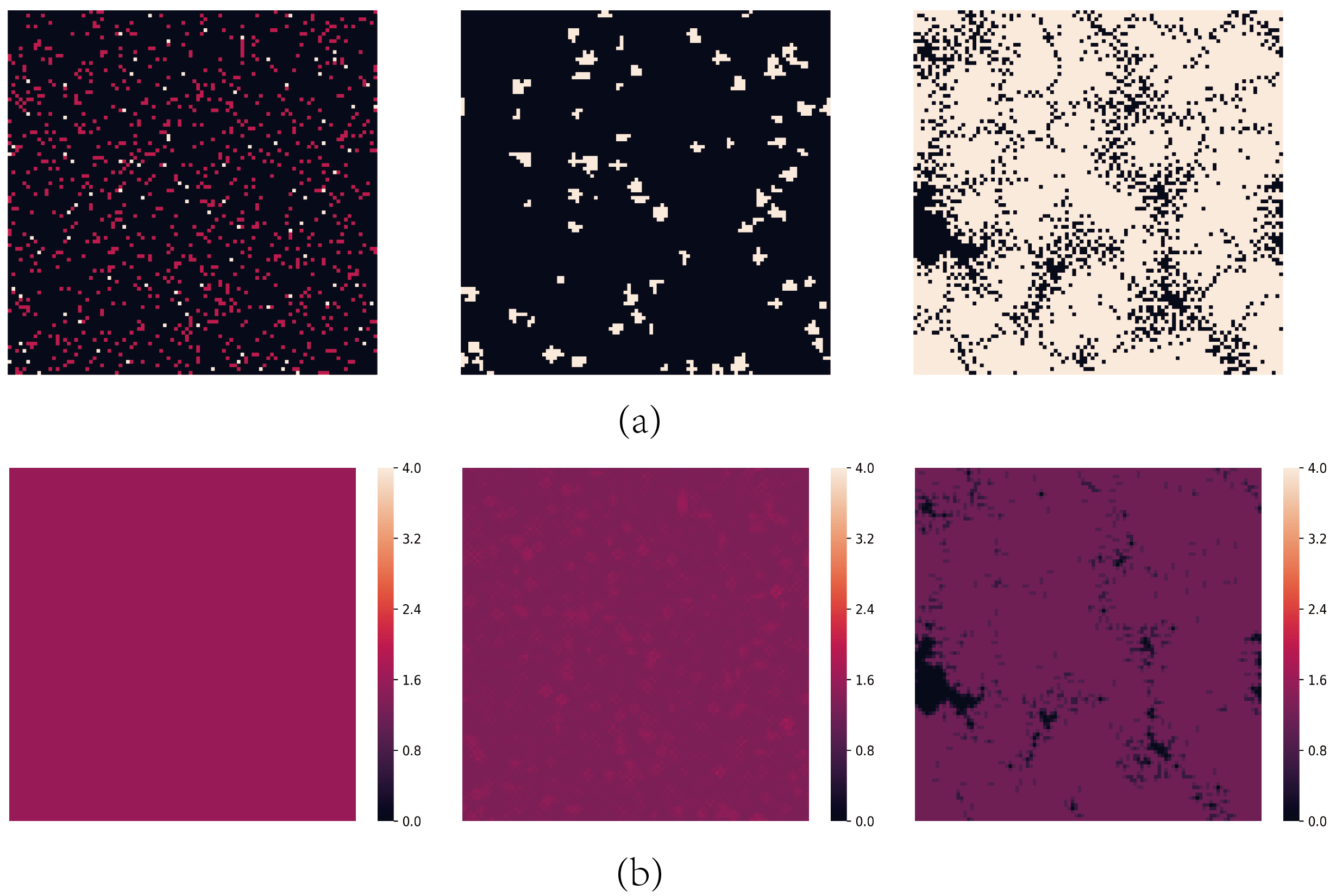}

\caption{(a) shows characteristic snapshots of cooperators(red), loners(white) and defectors(black) with time growing. (b) shows the heat map of aspiration distribution with time growing. The steps of them are $t$=0, 5 and 100 from left to right respectively. The results were got when $A=1.6$ and $b=1.6$ $r_{\mathcal{C}0}=0.09$, $r_{\mathcal{D}0}=0.9$ and $r_{\mathcal{L}0}=0.01$ initially.}\label{7}

\end{figure}

Besides, whether cooperators could survive is also worth considering.
Figure \ref{7} shows another adverse condition that cooperators will be extinct.
At first, nearly all cooperators are surrounded by defectors and loners.
They are dissatisfied and get lower payoffs than their $\mathcal{D}$ and $\mathcal{L}$ neighbors, so cooperators are extinct more quickly than loners.
Then loners could expand easily among defectors because a loner's payoff will always be larger than a defector's payoff if no cooperator exists.
As shown in Figure \ref{7} (a), loners will occupy most of the network when stable, but no cooperator exists.

\begin{figure}[!htpb]

\centering

\includegraphics[width=6in]{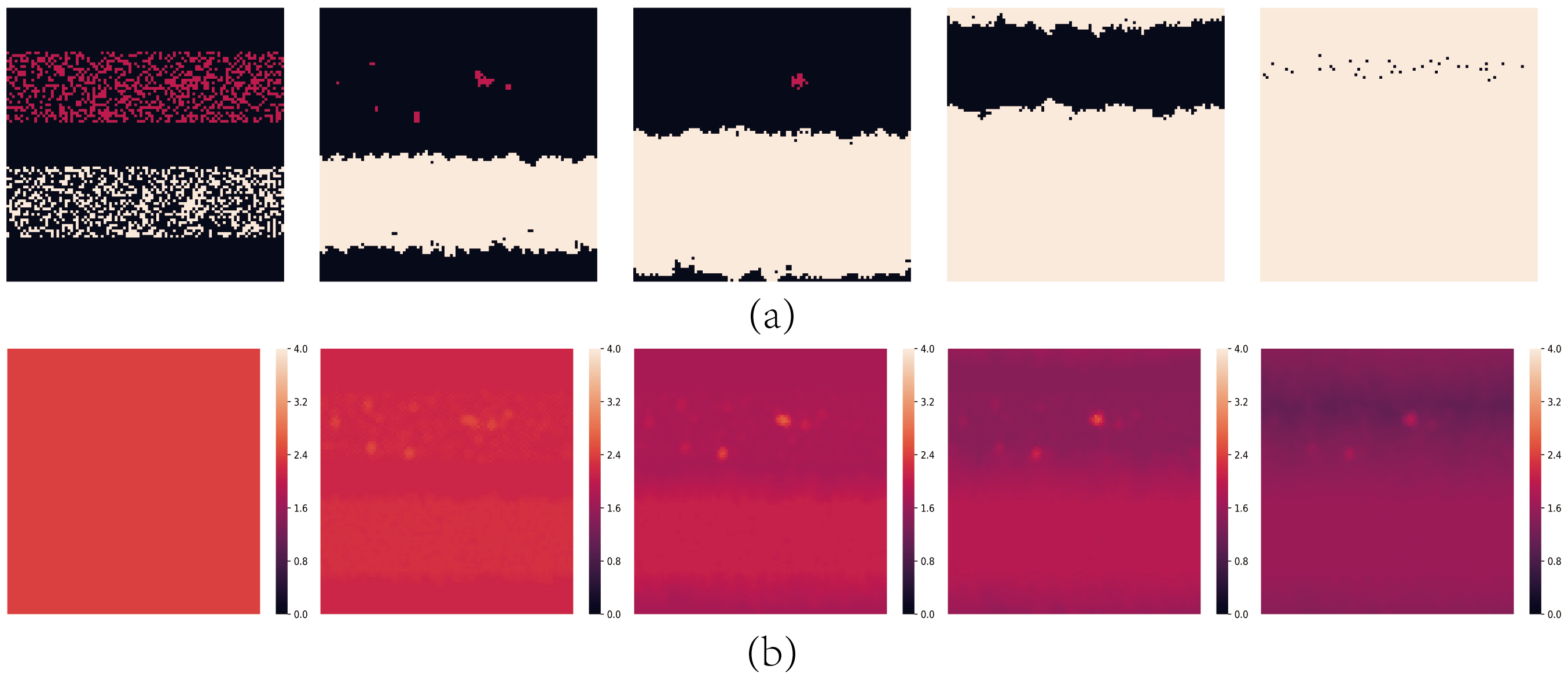}

\caption{(a) shows characteristic snapshots of cooperators(red), loners(white) and defectors(black) with time growing. (b) shows the heat map of aspiration distribution with time growing. The steps of them are $t$=0, 10, 30, 50 and 100 from left to right respectively. The results were got when $A=2.4$ and $b=1.6$ with $r_{\mathcal{C}0}=0.1$, $r_{\mathcal{D}0}=0.8$ and $r_{\mathcal{L}0}=0.1$ initially, where cooperators and loners are separated by defectors.}\label{8}

\end{figure}

\begin{figure}[!htpb]

\centering

\includegraphics[width=6in]{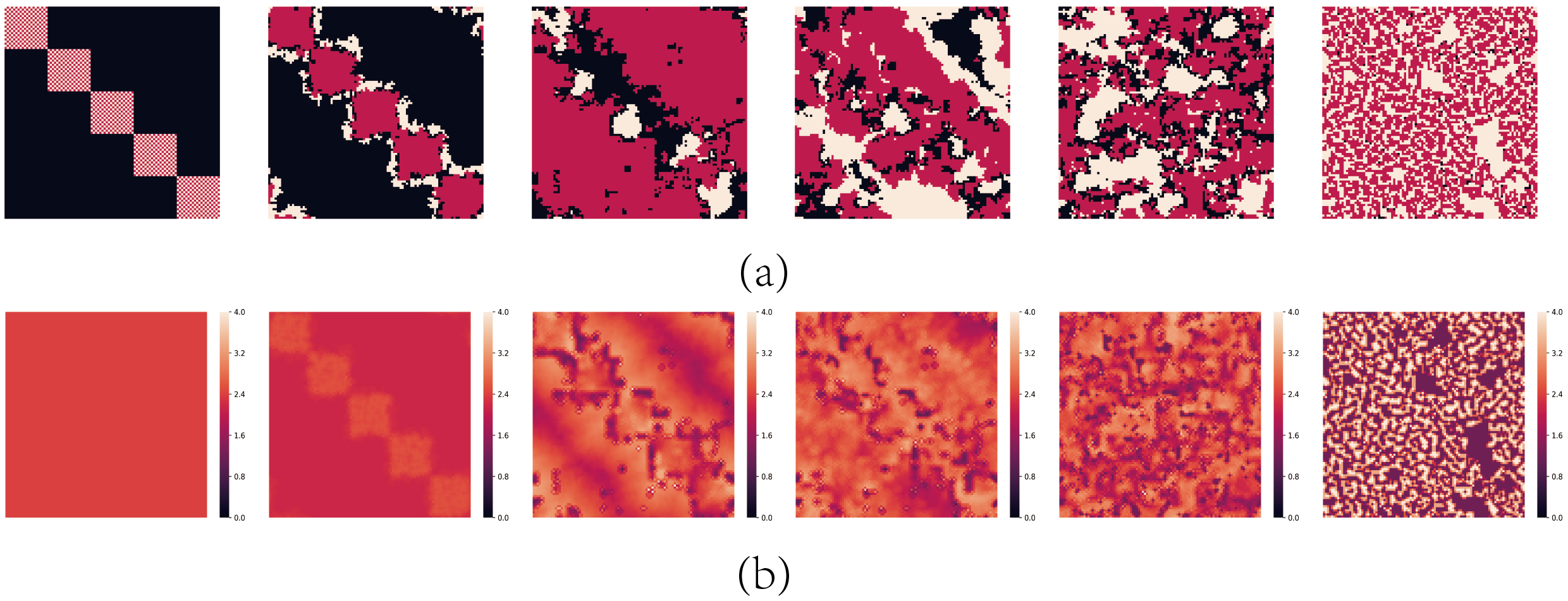}

\caption{(a) shows characteristic snapshots of cooperators(red), loners(white) and defectors(black) with time growing. (b) shows the heat map of aspiration distribution with time growing. The steps of them are $t$=0, 10, 100, 500, 1000 and 10000 from left to right respectively. The results were got when $A=2.4$ and $b=1.6$ with $r_{\mathcal{C}0}=0.1$, $r_{\mathcal{D}0}=0.8$ and $r_{\mathcal{L}0}=0.1$ initially, where cooperators and loners are mixed well.}\label{9}

\end{figure}

From the above, voluntary participation could promote cooperation only when both loners and cooperators are not extinct during the evolution process.
To further understand this condition, some special setup should be considered.
Figure \ref{8} shows the a typical special setup in which cooperators and loners are separated by defectors.
It is shown from Figure \ref{8} that in the upper half of the network, all cooperators are surrounded by defectors and they couldn't survive when $b=1.6$.
Meanwhile in the lower half of the network, loners expand because of the higher payoffs.
But when loners expand into the upper half of the network, cooperators have become extinct, so the loners will finally occupy the whole network and cooperation is not promoted.
Figure \ref{9} shows another initial setting for the same parameters and fractions of three strategies with Figure \ref{8}.
Cooperators and loners mix well in five clusters initially, and it could be observed that cooperators expand on the border of the clusters fast by the influence of loners. When loners form clusters and get satisfied, they will be stable, then cooperators and loners will coexist with moderate fractions.
Well-mixed cooperators and loners could expand together, which is the main reason why voluntary participation could promote cooperation.

To conclude, for small values of $A$ ($A \leq 1.2$), three kinds of strategies could coexist.
Loners will never change their strategies because of their fixed payoffs, while part of cooperators and defectors are dissatisfied and evolve into loners, which depends on the value of $A$ but is independent with the value of $b$.
Then the cyclic rock-scissors-paper type of dominance happens and three strategies could coexist.
For large values of $A$ ($A \textgreater 1.2$), cooperators and loners could expand and coexist while defectors' survival is greatly suppressed.
Loners could form some large stable clusters where all individuals are always satisfied so they could expand unconditionally.
Most of the defectors are hard to survive when there are a large number of loners in the network because they have no chance to get the temptation value $T$, even when it is large.
On the contrary, cooperators could coexist with loners easily by forming many small clusters.
Besides, it should be noticed that initial distribution has a significant impact on promoting cooperation.
Cooperators or loners may be extinct under the adverse initial distribution.
Loners should be fully adjacent to both cooperators and defectors, which is a necessary condition for promoting cooperation.
In traditional fixed aspiration model, voluntary participation plays a rather limited role in promoting cooperation under high aspiration
levels, where loners are always dissatisfied so they are hard to survive.
Defection cannot be suppressed when $b$ is large.
However, dynamic aspiration model provides a favorable environment for loners to survive and expand, an individual' aspiration could decrease to a low level if they get dissatisfied for a long term.
When an individual' aspiration becomes lower than 1.2, it could keep $\mathcal{L}$ as its strategy because of the fixed payoff.

\section{Conclusion}\label{section 5}
In summary, this paper discusses how voluntary participation impacts PD games with Dynamic-Win-Stay-Lose-Learn strategy updating rule.
This dynamic model is adjusted by a single parameter $a$.
It is found that the proposal of strategy $\mathcal{L}$ could promote cooperation and suppress defection within a wide range of parameters, especially when $b$ is large.
We also studied how initial distribution influences the evolutionary process to reveal some adverse initial distribution.
In OPD games, the best way for loners and cooperators to survive is mixing well and forming clusters gradually, which could be easily achieved with Dynamic-Win-Stay-Lose-Learn strategy updating rule.

Our work combine voluntary participation with dynamic aspiration model to provide a new perspective on how voluntary participation promote cooperation in PD games.
In terms of the broader relevance of our research, comparing with fixed aspiration model, dynamic aspiration model is more in line with the law of evolutionary games.
Dynamic aspiration is the natural tendency among humans, so it is expected that our work provides some reference values for solving the social dilemma in the real world \cite{wang2017real, wang2016real}.

\section*{Acknowledgments}
This work is supported by the Fundamental Research Funds for the Central Universities, the Research and Development Program of China (No.2018AAA0101100), the Beijing
Natural Science Foundation (1192012, Z180005) and National Natural Science Foundation of China (No.62050132).

\section*{Reference}

\bibliographystyle{unsrt}
\bibliography{ref}

\begin{thebibliography}{10}

\bibitem{Myerson1991Game}
Roger~B. Myerson.
\newblock {\em Game Theory: Analysis of Conflict}.
\newblock Harvard University Press, 1991.

\bibitem{schuster2004Game}
Richard Schuster and Amir Perelberg.
\newblock Why cooperate?: an economic perspective is not enough.
\newblock {\em Behavioural Processes}, 66(3):261--277, 2004.

\bibitem{gibbons1992Game}
Robert~S Gibbons.
\newblock {\em Game theory for applied economists}.
\newblock Princeton University Press, 1992.

\bibitem{axelrod1981evolution}
Robert Axelrod and William~Donald Hamilton.
\newblock The evolution of cooperation.
\newblock {\em science}, 211(4489):1390--1396, 1981.

\bibitem{nowak2006five}
Martin~A Nowak.
\newblock Five rules for the evolution of cooperation.
\newblock {\em science}, 314(5805):1560--1563, 2006.

\bibitem{nowak1992spatial}
Martin~A Nowak and Robert~M May.
\newblock Evolutionary games and spatial chaos.
\newblock {\em Nature}, 359(6398):826--829, 1992.

\bibitem{Szab1998lattice}
Gy?Rgy Szab¨® and Csaba T?Ke.
\newblock Evolutionary prisoner's dilemma game on a square lattice.
\newblock {\em Physical Review E}, 58(1):69--73, 1998.

\bibitem{2005smallworld}
Zx~Wu, Xj~Xu, Y~Chen, and Yh~Wang.
\newblock Spatial prisoner's dilemma game with volunteering in newman-watts
  small-world networks.
\newblock {\em Physical Review E Statistical Nonlinear \& Soft Matter Physics},
  71(3):037103, 2005.

\bibitem{2007smallworld}
Jie Ren, Wen~Xu Wang, and Feng Qi.
\newblock Randomness enhances cooperation: A resonance-type phenomenon in
  evolutionary games.
\newblock {\em Physical Review E}, 75(4 Pt 2):045101, 2007.

\bibitem{2007scalefree}
Zhihai Rong, Xiang Li, and Xiaofan Wang.
\newblock Roles of mixing patterns in cooperation on a scale-free networked
  game.
\newblock {\em Physical Review E Statistical Nonlinear \& Soft Matter Physics},
  76(2):027101, 2007.

\bibitem{Du2009scalefree}
Wen~Bo Du, Xian~Bin Cao, Zhao Lin, and Mao~Bin Hu.
\newblock Evolutionary games on scale-free networks with a preferential
  selection mechanism.
\newblock {\em Physica A-statistical Mechanics \& Its Applications},
  388(20):4509--4514, 2009.

\bibitem{herrmann2008punishment}
Benedikt Herrmann, Christian Th{\"o}ni, and Simon G{\"a}chter.
\newblock Antisocial punishment across societies.
\newblock {\em Science}, 319(5868):1362--1367, 2008.

\bibitem{helbing2010punishment}
Dirk Helbing, Attila Szolnoki, Matja{\v{z}} Perc, and Gy{\"o}rgy Szab{\'o}.
\newblock Punish, but not too hard: how costly punishment spreads in the
  spatial public goods game.
\newblock {\em New Journal of Physics}, 12(8):083005, 2010.

\bibitem{chen2014punishment}
Xiaojie Chen, Attila Szolnoki, and Matja{\v{z}} Perc.
\newblock Probabilistic sharing solves the problem of costly punishment.
\newblock {\em New Journal of Physics}, 16(8):083016, 2014.

\bibitem{chen2015punishment}
Xiaojie Chen, Tatsuya Sasaki, {\AA}ke Br{\"a}nnstr{\"o}m, and Ulf Dieckmann.
\newblock First carrot, then stick: how the adaptive hybridization of
  incentives promotes cooperation.
\newblock {\em Journal of the royal society interface}, 12(102):20140935, 2015.

\bibitem{szolnoki2015conformist}
Attila Szolnoki and Matja{\v{z}} Perc.
\newblock Conformity enhances network reciprocity in evolutionary social
  dilemmas.
\newblock {\em Journal of The Royal Society Interface}, 12(103):20141299, 2015.

\bibitem{szolnoki2016conformist}
Attila Szolnoki and Matja{\v{z}} Perc.
\newblock Leaders should not be conformists in evolutionary social dilemmas.
\newblock {\em Scientific Reports}, 6:23633, 2016.

\bibitem{2018compassion}
Yumeng Li, Jun Zhang, and Matja Perc.
\newblock Effects of compassion on the evolution of cooperation in spatial
  social dilemmas.
\newblock {\em Applied Mathematics and Computation}, 320:437--443, 2018.

\bibitem{2006Memory}
Wen~Xu Wang, Jie Ren, Guanrong Chen, and Bing~Hong Wang.
\newblock Memory-based snowdrift game on networks.
\newblock {\em Physical Review E}, 74(5 Pt 2):056113, 2006.

\bibitem{2008Memory}
Sm~Qin, Y~Chen, Xy~Zhao, and J~Shi.
\newblock Effect of memory on the prisoner's dilemma game in a square lattice.
\newblock {\em Physical Review E Statistical Nonlinear \& Soft Matter Physics},
  78(4), 2008.

\bibitem{Yang2012aspiration}
Han~Xin Yang, Zhihai Rong, Pei~Min Lu, and Yong~Zhi Zeng.
\newblock Effects of aspiration on public cooperation in structured
  populations.
\newblock {\em Physica A Statal Mechanics \& Its Applications},
  391(15):4043--4049, 2012.

\bibitem{Wu2018aspiration}
Te~Wu, Feng Fu, and Long Wang.
\newblock Coevolutionary dynamics of aspiration and strategy in spatial
  repeated public goods games.
\newblock {\em New Journal of Physics}, 20(6), 2018.

\bibitem{Chu2019aspiration}
Chen Chu, Chunjiang Mu, Jinzhuo Liu, Chen Liu, Stefano Boccaletti, Lei Shi, and
  Zhen Wang.
\newblock Aspiration-based coevolution of node weights promotes cooperation in
  the spatial prisoner's dilemma game.
\newblock {\em New Journal of Physics}, 2019.

\bibitem{Zhang2019aspiration}
Liming Zhang, Changwei Huang, Haihong Li, and Qionglin Dai.
\newblock Aspiration-dependent strategy persistence promotes cooperation in
  spatial prisoner's dilemma game.
\newblock {\em Epl}, 126(1):18001, 2019.

\bibitem{posch1999dynamic}
Martin Posch, Alexander Pichler, and Karl Sigmund.
\newblock The efficiency of adapting aspiration levels.
\newblock {\em Proceedings of the Royal Society of London. Series B: Biological
  Sciences}, 266(1427):1427--1435, 1999.

\bibitem{amaral2016dynamic}
Marco~A Amaral, Lucas Wardil, Matja{\v{z}} Perc, and Jafferson~KL da~Silva.
\newblock Stochastic win-stay-lose-shift strategy with dynamic aspirations in
  evolutionary social dilemmas.
\newblock {\em Physical Review E}, 94(3):032317, 2016.

\bibitem{arefin2020dynamic}
Md~Rajib Arefin and Jun Tanimoto.
\newblock Evolution of cooperation in social dilemmas under the coexistence of
  aspiration and imitation mechanisms.
\newblock {\em Physical Review E}, 102(3):032120, 2020.

\bibitem{2020dynamic}
Cong Li and Suohai Fan.
\newblock A dynamic aspiration-based interaction strategy blocks the spread of
  defections in social dilemma.
\newblock {\em EPL (Europhysics Letters)}, 129(4):48002 (7pp), 2020.

\bibitem{shi2021dynamic}
Zhenyu Shi, Wei Wei, Xiangnan Feng, Xing Li, and Zhiming Zheng.
\newblock Dynamic aspiration based on win-stay-lose-learn rule in spatial
  prisoner¡¯s dilemma game.
\newblock {\em Plos one}, 16(1):e0244814, 2021.

\bibitem{liu2012win}
Yongkui Liu, Xiaojie Chen, Lin Zhang, Long Wang, and Matja{\v{z}} Perc.
\newblock Win-stay-lose-learn promotes cooperation in the spatial prisoner's
  dilemma game.
\newblock {\em PloS one}, 7(2):e30689, 2012.

\bibitem{chu2017win}
Chen Chu, Jinzhuo Liu, Chen Shen, Jiahua Jin, and Lei Shi.
\newblock Win-stay-lose-learn promotes cooperation in the prisoner's dilemma
  game with voluntary participation.
\newblock {\em Plos one}, 12(2):e0171680, 2017.

\bibitem{fu2018win}
Ming-Jian Fu and Han-Xin Yang.
\newblock Stochastic win-stay-lose-learn promotes cooperation in the spatial
  public goods game.
\newblock {\em International Journal of Modern Physics C}, 29(04):1850034,
  2018.

\bibitem{Szolnoki2009tit}
Attila Szolnoki, Matja Perc, and Gy?Rgy Szab¨®.
\newblock Phase diagrams for three-strategy evolutionary prisoner's dilemma
  games on regular graphs.
\newblock {\em Physical Review E}, 80(5):056104, 2009.

\bibitem{2017punishment}
Attila Szolnoki and Xiaojie Chen.
\newblock Alliance formation with exclusion in the spatial public goods game.
\newblock {\em Phys.rev.e}, 95(5-1):052316, 2017.

\bibitem{2016lone}
Marcos Cardinot, Colm O'Riordan, and Josephine Griffith.
\newblock The optional prisoner's dilemma in a spatial environment: Coevolving
  game strategy and link weights.
\newblock In {\em 8th International Conference on Evolutionary Computation
  Theory and Applications}, 2016.

\bibitem{szabo1998noise}
Gy{\"o}rgy Szab{\'o} and Csaba T{\H{o}}ke.
\newblock Evolutionary prisoner's dilemma game on a square lattice.
\newblock {\em Physical Review E}, 58(1):69, 1998.

\bibitem{szabo2005noise}
Gy{\"o}rgy Szab{\'o}, Jeromos Vukov, and Attila Szolnoki.
\newblock Phase diagrams for an evolutionary prisoner's dilemma game on
  two-dimensional lattices.
\newblock {\em Physical Review E}, 72(4):047107, 2005.

\bibitem{perc2006noise}
Matja{\v{z}} Perc.
\newblock Coherence resonance in a spatial prisoner's dilemma game.
\newblock {\em New Journal of Physics}, 8(2):22, 2006.

\bibitem{ogasawara2014end}
Takashi Ogasawara, Jun Tanimoto, Eriko Fukuda, Aya Hagishima, and Naoki
  Ikegaya.
\newblock Effect of a large gaming neighborhood and a strategy adaptation
  neighborhood for bolstering network reciprocity in a prisoner's dilemma game.
\newblock {\em Journal of Statistical Mechanics: Theory and Experiment},
  2014(12):P12024, 2014.

\bibitem{kabir2018end}
KM~Ariful Kabir, Jun Tanimoto, and Zhen Wang.
\newblock Influence of bolstering network reciprocity in the evolutionary
  spatial prisoner's dilemma game: A perspective.
\newblock {\em The European Physical Journal B}, 91(12):312, 2018.

\bibitem{wang2017real}
Zhen Wang, Marko Jusup, Rui-Wu Wang, Lei Shi, Yoh Iwasa, Yamir Moreno, and
  J{\"u}rgen Kurths.
\newblock Onymity promotes cooperation in social dilemma experiments.
\newblock {\em Science advances}, 3(3):e1601444, 2017.

\bibitem{wang2016real}
Zhen Wang, Chris~T Bauch, Samit Bhattacharyya, Alberto d'Onofrio, Piero
  Manfredi, Matja{\v{z}} Perc, Nicola Perra, Marcel Salath{\'e}, and Dawei
  Zhao.
\newblock Statistical physics of vaccination.
\newblock {\em Physics Reports}, 664:1--113, 2016.

\end{thebibliography}


\end{document}